\documentclass[letterpaper,pra,twocolumn,showpacs,superscriptaddress]{revtex4}
\usepackage{graphicx}
\usepackage{amsmath}
\usepackage{amssymb}

\begin{document}

\title{Optical storage of high density information beyond the diffraction limit : a quantum study}
\author{V. Delaubert}
\affiliation{Laboratoire Kastler Brossel, UPMC, Case 74, 4 Place Jussieu, 75252 Paris cedex 05, France}
\author{N. Treps}
\affiliation{Laboratoire Kastler Brossel, UPMC, Case 74, 4 Place Jussieu, 75252 Paris cedex 05, France}
\author{G. Bo}
\affiliation{Laboratoire Pierre Aigrain, Ecole Normale Sup\'{e}rieure, 24 rue Lhomond, 75231 Paris Cedex 05, France}
\author{C. Fabre}
\affiliation{Laboratoire Kastler Brossel, UPMC, Case 74, 4 Place Jussieu, 75252 Paris cedex 05, France}

\date{\today}

\begin{abstract}
We propose an optical read-out scheme allowing a demonstration of principle of information extraction below the
diffraction limit. This technique, which could lead to improvement in data read-out density onto optical discs, is
independent from the wavelength and numerical aperture of the reading apparatus, and involves a multi-pixel array
detector. Furthermore, we show how to use non classical light in order to perform bit discrimination beyond the quantum
noise limit.
\end{abstract}


\maketitle

\section*{Introduction}

The reconstruction of an object from its image beyond the diffraction limit, typically of the order of the wavelength, is
a hot field of research, though a very old one, as Bethe already dealt with the theory of diffraction by sub-wavelength
holes in 1944 \cite{Bethe}. More recently, theory has been developed to be applied to the optical storage problem, in
order to study the influence of very small variations of pit width or depth relative to the wavelength \cite{Bethe, Marx1,
Marx2, Wang, Liu, Brok}. To date, only a few super-resolution techniques \cite{Kolobov} include a quantum treatment of the
noise in the measurement, but  to our knowledge, none has been applied to the optical data storage problem.

Optical discs are now reaching their third generation, and have improved their data capacity from 0.65~GB for compact
discs (using a wavelength of 780~nm), to 4.7~GB for DVDs ($\lambda$ =~650~nm), and eventually to 25~GB for the Blu-Ray
discs (using a wavelength of 405~nm). In addition to new coding techniques, this has been achieved by reducing the spot
size of the diffraction-limited focused laser beam onto the disc, involving higher numerical apertures and shorter
wavelengths.

Several further developments are now in progress, such as the use of volume holography, 266~nm reading lasers, immersion
lenses, near field systems, multi-depths pits \cite{Hsu}, or information encoding on angle positions of asymmetrical pits
\cite{TorokCD}. These new techniques rely on bit discrimination using small variations of the measured signals. Therefore,
the noise is an important issue, and ultimately, quantum noise will be the limiting factor.

In this paper, we investigate an alternative and complementary way to increase the capacity of optical storage, involving
the retrieval of information encoded on a scale smaller than the wavelength of the optical reading device. We investigate
a way to optimize the detection of sub-wavelength structures using multi-pixel array. An attempt to a full treatment of
the optical disc problem being far too complex, we have chosen to illustrate our proposal on a very simple example,
leaving aside most technical constraints and complications, but still involving all the essence of the overall problem.

We first explain how the use of an array detector can lead to an improvement of the detection and distinction of
sub-wavelength structures present in the focal spot of a laser beam. We then focus on information extraction from an
optical disc with a simple but illustrative example, considering that only a few bits are burnt on the dimensions of the
focal spot of the reading laser, and show how the information is encoded from the disc to the light beam, propagated to
the detector, and finally detected. We explain the gain configuration of the array detector that has to be chosen in order
to improve the signal-to-noise ratio (SNR) of the detection. Moreover, as quantum noise is experimentally accessible, and
will be a limiting factor for further improvements, we perform a quantum calculation of the noise in the detection
process. Indeed, we present how this detection can be optimized to perform simultaneous measurements below the quantum
noise limit, using non classical light.

\section{Proposed scheme for bit sequence recognition in optical discs}

\begin{figure}[htbp]
\begin{center}
\includegraphics[width=7cm]{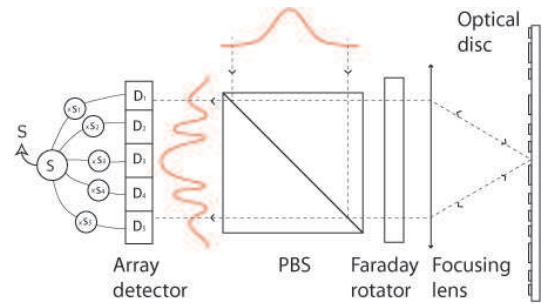}
\caption{Color 
online {\bf Scheme for information extraction from optical disc, using an array detector.}} \label{general_scheme}
\end{center}
\end{figure}
We propose a novel optical read-out scheme shown on figure (\ref{general_scheme}) allowing information extraction from
optical discs beyond the diffraction limit, based on multi-pixel detection. Bits, coded as pits and holes on the optical
disc, induce phase flips in the electric field transverse profile of the incident beam at reflection. The reflected beam
is imaged in the far field of the disc plane, where the detector stands. In the far field, the phase profile induced by
the disc is converted into an intensity profile, that the multi-pixel detectors can, at least partly, reconstruct.

Taking into account that a lot of a priori information is available - i.e. only a finite number of intensity profiles is
possible - we propose to use a detector with a limited number of pixels $D_{k}$ whose gains can independently be varied
depending on which bit sequence one wants to detect. The signal is then given by
\begin{eqnarray}\label{signal}
S = \sum_{k}\sigma_{k}N_{k}
\end{eqnarray}
where $N_{k}$ is the mean photon number detected on pixel $D_{k}$, and $\sigma_{k}$ is the electronic gain of the same
pixel. Ideally, to each bit sequence present on the disc corresponds a set of gains chosen so that the value of the
measurement is zero, thus cancelling noise from the mean field. Measuring the signal for a given time interval $T$ around
the centered position of a bit sequence in the focal spot, and testing, in parallel, all the pre-define sets of gain in
the remaining time, allows to deduce which bit sequence is present on the disc.

We will first show that this improvement in density of information
encoded on an optical disc is already possible using classical
resources. Moreover, as the measurement is made around a zero mean
value, the classical noise is mostly cancelled. Hence, we reach
regimes where the quantum noise can be the limiting factor. We
will demonstrate how to perform measurements beyond the quantum
noise limit, using previous results on quantum noise analysis in
multi-pixel detection developed in reference \cite{Treps}.

\section{Encoding information from a disc onto a light beam}

We have explained the general principle of reading-out sub-wavelength bit sequences encoded on an optical disc, and now
focus on the information transfer from the optical disc to the laser beam, through an illustrative example.

\begin{figure}[htbp]
\begin{center}
\includegraphics[width=7cm]{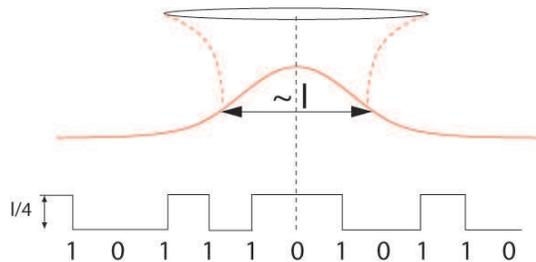}
\caption{Color 
online {\bf Example of bit sequence on an optical disc. The spacing between the bits is smaller than the wavelength, the
minimum waist of the incident laser beam being of the order of $\lambda$. A hole depth of $\lambda/4$ insures a $\pi$
phase shift between fields reflected on a pit and a hole.}} \label{bits}
\end{center}
\end{figure}

Let us recall that bits are encoded by pits and holes on the disc surface: a step change from hole to pit (or either from
pit to hole) encodes bit $1$, whereas no depth change on the surface encodes bit $0$, as represented on figure
(\ref{bits}). A hole depth of $\lambda/4$ insures a $\pi$ phase shift between the fields reflected by a pit and a hole. In
this section, we compute the incident field distribution on the optical disc affected by the presence of a bit sequence in
the focal spot, and finally analyze the intensity back reflected in the far field, in the detection plane, as sketched on
figure (\ref{general_scheme}).

\subsection{Beam focalization}

Current optical disc read-out devices involve a linearly polarized beam strongly focused on the disc surface to point out
details whose size is of the order of the laser wavelength. The numerical aperture ($NA$) of the focusing lens can be
large ($0.47$ for CDs, $0.6$ for DVDs, and $0.85$ for BLU RAY discs), and the exact calculation of the field cannot be
done in the paraxial and scalar approximation. Thus, the vectorial theory of diffraction has to be taken into account.

The structure of the electromagnetic field in the focal plane of a
strongly focused beam has been investigated for decades now
\cite{VanNie}, as its applications include areas such as
microscopy, laser micro-fabrication, micromanipulation, and
optical storage \cite{Landesman, Rodriguez, Ulanowski, Lax,
Seshadri, Ciattoni, Cao, Nieminen}.

\begin{figure}[htbp]
\begin{center}
\includegraphics[width=9cm]{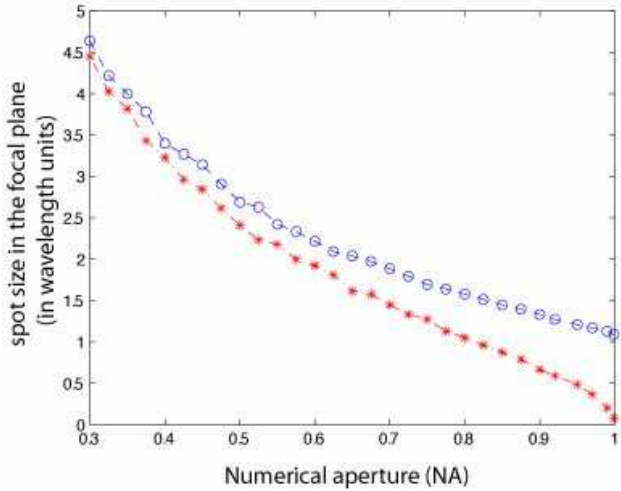}
\caption{Color 
online{\bf Evolution of the focused spot size of an incident plane wave with the numerical aperture (for
$\lambda$~=~780 nm in air medium). The spot size is limited to the order of the wavelength in the non-paraxial case (o),
whereas it goes to zero for very high numerical apertures in the paraxial case($\ast$).}} \label{waistcomparison}
\end{center}
\end{figure}

In our case of interest, we can restrict the field calculation to
the focal plane, which is the disc plane. Thus Richards and Wolf
integrals \cite{Richards}, which are not suitable for a general
propagation of the field, but which can provide the field profile
in the focal plane for any type of polarization of the incoming
beam as long as the focusing length is much larger than the
wavelength, can be used to achieve this calculation. These
integrals have already been used in many publications dealing with
tight focusing processes \cite{Quabis1, Novotny, Dorn, Quabis2,
Sheppard, Torok, Youngworth, Zhan}. As highlighted in these
references, the importance of the vectorial aspect of the field
can easily be understood when a linearly polarized beam is
strongly focused, as the polarization of the wave after the lens
is not perpendicular to the propagation axis anymore and has thus
components along this axis. In order to estimate the limit of
validity of the paraxial approximation, we computed focused spot
sizes of linearly polarized beam in the focal plane for different
numerical apertures, first in the paraxial approximation, and then
calculated with Richards and Wolf integrals. The results are
compared on figure (\ref{waistcomparison}) for an incident plane
wave in air medium with $\lambda=780nm$, where the spot size is
defined as the diameter which contains $86\%$ of the focused
energy, as in reference \cite{Siegman}.
We see that when the numerical aperture exceeds $0.6$, a good prediction requires a non-paraxial treatment. Moreover,
whereas there is no theoretical limit to focalization in the paraxial case, we see that non paraxial effects prevent us to
reach a waist smaller than the order of the wavelength. Note that this limit is not fundamental and can be overcome by
modifying the polarization of the incoming beam. Quabis {\it et al.} have indeed managed to reduce the spot area to about
$0.1~\lambda^{2}$ using an incident radially polarized doughnut beam \cite{Quabis1, Quabis2}.

\begin{figure}[htbp]
\begin{center}
\includegraphics[width=7cm]{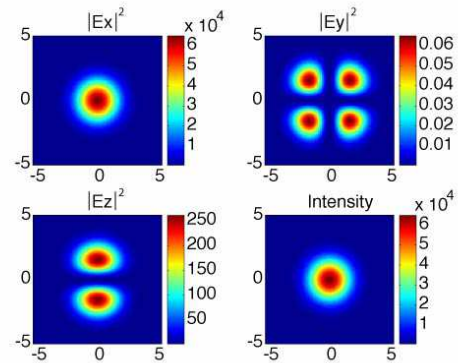}
\caption{Color 
online{\bf Norm of the different field components and resultant intensity in the focal plane with a linearly polarized
incident field along the $x$ axis, focused with a $0.47$ numerical aperture.}} \label{fieldcomponents}
\end{center}
\end{figure}

As our aim is to present a demonstration of principle and not a full treatment of the optical disc problem, the following
calculations will be done using the physical parameters of the actual Compact Discs ($\lambda=780nm$ and $NA=0.47$,
corresponding to a focalization angle of $27$ degrees in air medium). In this case, the paraxial and scalar approximations
are still valid. Indeed, figure (\ref{fieldcomponents}), giving the transverse profile of the three field components and
the resultant intensity in the focal plane using the former parameters, shows that although the field is not strictly
linearly polarized as foreseen before, $E_{y}\ll E_{z}\ll E_{x}$, and we can thus consider that only $E_{x}$ is different
from zero with a good approximation. Note that the exact expression would not intrinsically change the problem, as our
scheme can be adapted to any field profile discrimination.

\subsection{Reflection onto the disc}

In order to compute the reflected field, we simply assume that bumps and holes are generated in such a way that they
induce a $\pi$ phase shift between them at reflection on the field profile. Note that the holes depth is usually
$\lambda/4$, but precise calculations would be required to give the exact shape of the pits, as they are supposed to be
burnt below the wavelength size, and as the field penetration in those holes is not trivial \cite{Wang, Brok, Liu}. As we
have shown that only one vectorial component of the field was relevant in the focal plane, we can directly apply this
phase shift to the amplitude profile of this component.

We first envision a scheme with only three bits in the focal spot, which means that $2^{3}$ different bit sequences, i.e.
a byte, have to be distinguished from each other, using the information extracted from the reflected field. Note that we
neglect the influence of other bits in the neighborhood. A more complete calculation involving this effect with more bits
will be considered in a further approach.

\begin{figure}[htbp]
\begin{center}
\includegraphics[width=9cm]{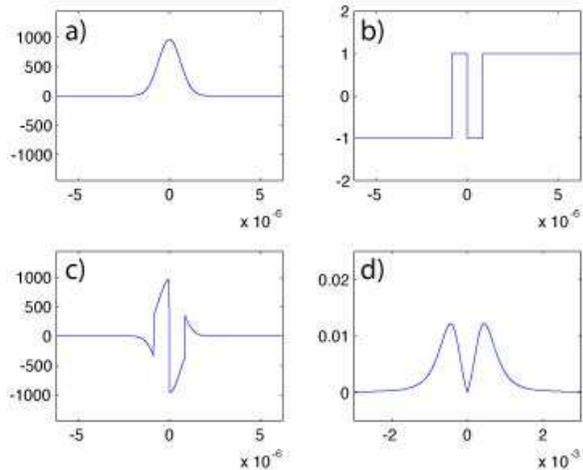}
\caption{Color 
online{\bf Modifications of the transverse amplitude field profile trough propagation, in the case of a $111$ bit
sequence in the focal spot : a) incoming beam profile, b) 111 bit sequence, c) corresponding reflected field in the disc
plane, d) far field profile in the detector plane .}} \label{summary}
\end{center}
\end{figure}

The amplitude profiles obtained when the incident beam is centered on a bit of the CD are presented on figure
(\ref{summary}), for a particular bit sequence. Note that we have chosen the space between two bits on the disc equal to
the waist size of the reading beam. The first three curves respectively show the field amplitude profile incident on the
disc, an example of a bit sequence, and the corresponding profile just after reflection onto the disc. We see that binary
information is encoded from bumps and holes on the CD to phase flips in the reflected field.

\subsection{Back propagation to the detector plane}

In order to extract the information encoded in the transverse amplitude profile of the beam, the field has to be back
propagated to the detector plane. A circulator, composed of a polarizing beam splitter and a Faraday rotator, ensures that
the linearly polarized reflected beam reaches the array detector, as shown on figure (\ref{general_scheme}). Assuming that
the detector is positioned just behind the lens plane, the expression of the detected field is given by the far field of
the disc plane, apertured by the diameter of the focusing lens. As the focal length and the diameter of the lens are large
compared to the wavelength, we use Rayleigh Sommerfeld integral to compute the field in the lens plane \cite{Born}. As an
example, the calculated far field profile when the bit sequence $111$ is present in the focal spot is shown on the fourth
graph of figure (\ref{summary}).

The presence of the lens provides a limited aperture for the beam and cuts the high spatial frequencies of the field,
which can be a source of information loss, as the difference between each bit sequence can rely on those high frequencies.
However, we will see that enough information remains in the low frequency part of the spatial spectrum, so that the $8$
bits can be distinguished. This is due to the fact that we have in this problem a lot of a priori information on the
possible configurations to distinguish.

\begin{figure}[htbp]
\begin{center}
\includegraphics[width=9cm]{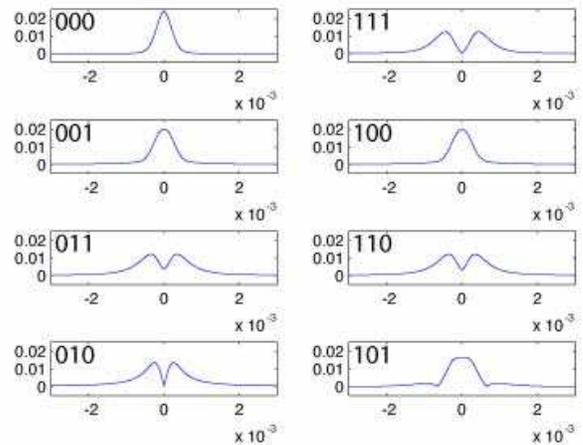}
\caption{Color 
online{\bf Field profiles in the array detector plane, for each of the $8$ bit sequence configuration. Note that they
are clearly distinguishable, except for the bit sequences $100$ and $001$, and $011$ and $110$, which have the same
profile because of the symmetry of the bit sequence relative to the position of the incident laser beam. }}
\label{far_field}
\end{center}
\end{figure}

\begin{figure}[htbp]
\begin{center}
\includegraphics[width=9cm]{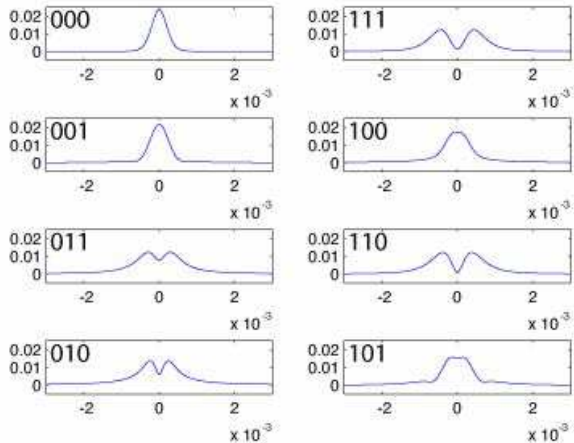}
\caption{Color 
online{\bf Field profiles in the array detector plane, for each of the $8$ bit sequence configuration, when the
position of the disc has been shifted of $w_{0}/6$ relative to the incident beam. The profile degeneracy for $100$ and
$001$, and $011$ and $110$ is raised. Note that the other profiles have experienced a much smaller shape modification
between the two positions of the disc. }} \label{non_centered}
\end{center}
\end{figure}

We see on figure (\ref{far_field}) that, with the physical parameters used in compact disc read out devices, 6 over 8
profiles in the detector plane are still different enough to be distinguished.
At this stage, we are nevertheless unable to discriminate between symmetric configurations, because they give rise to the
same far field profile. Therefore, 100 and 001, and 110 and 011, cannot be distinguished. Note that this problem can be
solved thanks to the rotation of the disc. Indeed, an asymmetry is created when the position of the disc relative to the
laser beam is shifted, thus modifying differently the two previously indistinguishable profiles. As shown on figure
(\ref{non_centered}), where the far field profiles are represented after a shift of $w_{0}/6$ in the position of the disc,
the degeneracy has been removed. Moreover, it is important to notice that the other profiles experience a small shape
modification. This redundant information is very useful in order to remove ambiguities while the disc is rotating.

\section{Information extraction for bit sequence recognition}

In this section, we describe the detection,present some
illustrative results, and the way they can be used to increase the
read-out precision of information encoded on optical discs. We
show here that a pixellised detector with a very small number of
pixels is enough to distinguish between the $8$ bit sequences.
Note that for technical and computing time reasons, it is not
realistic to use a CCD camera to record the reflected images, as
such cameras cannot yet combine good quantum efficiency and high
speed.

\subsection{Detected profiles}

\begin{figure}[htbp]
\begin{center}
\includegraphics[width=8cm]{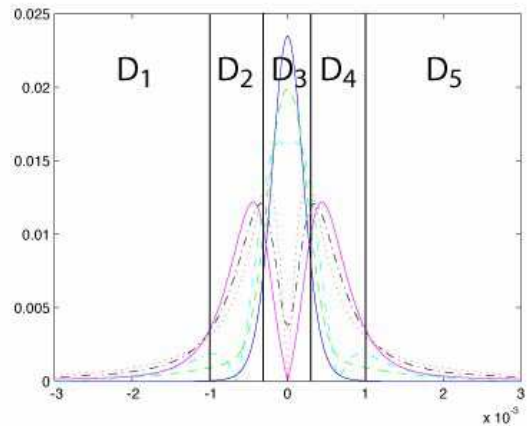}
\caption{Color 
online{\bf Far field profiles in detection the plane for each bit configuration, and array detector geometry. The 5
detectors $D_{1}..D_{5}$ have electronic gains $\sigma_{1}(i)..\sigma_{5}(i)$ according to the bit sequence $i$ which is
present in the focal spot.}} \label{FF_detection}
\end{center}
\end{figure}

For simplicity reason, we limit our calculation to a $5$ pixels array detector $D_{1}..D_{5}$, each of whom has an
electronic gain $\sigma_{1}..\sigma_{5}$, as shown on figure (\ref{FF_detection}). The size of each detector has been
chosen without a systematic optimization, which will be done in a further approach. Gain values are adapted to detect a
mean signal equal to zero for each bit configuration present in the focal spot, in order to cancel the common mode
classical noise present in the mean field \cite{Treps}. It means that for each bit sequence $i$, gains are chosen to
satisfy the following relation
\begin{equation}\label{gaindef}
\sum_{k=1}^{5} \sigma_{k}(i)N_{k}(i) =0
\end{equation}
where $N_{k}(i)$ is the mean photon number detected on pixel
$D_{k}$ when bit $i$ is present in the focal spot on the disc
\begin{equation}\label{Idef}
N_{k}(i) = \int_{D_{k}}n_{i}(x)dx
\end{equation}
where $n_{i}(x)$ is the number of photon incident on the array detector, at position $x$, when bit sequence $i$ is present
in the focal spot.

As all profiles are symmetrical when the incident beam is centered on a bit, we have set $\sigma_{1}=\sigma_{5}$ and
$\sigma_{2}=\sigma_{4}$. In addition, we have chosen $\sigma_{3}=-\frac{\sigma_{1}}{2}$. Using these relations and
equation (\ref{gaindef}), we compute gain values adapted to the recognition of each bit sequence.
Note that the calculation of each gain configuration requires a priori information on the far field profiles, or at least
an experimental calibration using a well-known sample.

Now that these gain configurations are set, we can investigate for a bit sequence on the optical disc.

\subsection{Classical results}

The expression of the detected signal $S_{i}(j)$ is given by
\begin{eqnarray}\label{signal}
S_{i}(j) = \sum_{k=1}^{5}\sigma_{k}(j)N_{k}(i)
\end{eqnarray}
where $i$ refers to the bit sequence effectively present in the focal spot, and $j$ to the gain set adapted to the
detection of the bit sequence $j$. It merely corresponds to the intensity weighted by the electronic gains. Note that for
$i=j$ - and only in this case if the detector is well chosen - the mean value of the signal $S_{i}(i)$ is equal to zero,
according to equation~(\ref{gaindef}). All possible values of $S_{i}(j)$ are presented for a total number of incident photons $N_{inc}=25$, in table (\ref{table}) where $i$
is read vertically, and corresponds to the bit sequence on the disc, whereas $j$ is read horizontally and refers to the
gain set adapted to the detection of bit $j$. In order not to have redundant information, we have gathered results
corresponding to identical far field profiles. A zero value is obtained for only one gain configuration, allowing an
identification of the bit sequence present in the focal spot.
\begin{table}
\begin{tabular}{|c|c|c|c|c|c|c|}

  \hline
          & 000 & 001/100 & 010  & 011/110 & 101   & 111 \\ \hline
  000     & 0   &-34 &-204 &   -254   &-77    &  -303 \\
  001/100 & 15  &0  & -76  &  -99   &-19  &  -121 \\
  010     & 23 &20 &0  & -6  &16 & -13 \\
  011/110 & 24 &22 &5  &0 & 19  & -5\\
  101     &  19& 11 & -36 &   -50   & 0 &-63\\
  111     & 24 &23 &9 &5   &20   &0 \\ \hline
\end{tabular}
\caption{{\bf Detected signals $S_{i}(j)$ where $i$ is read vertically and corresponds to the bit sequence on the disc,
whereas $j$ is read horizontally and refers to the gain set adapted to the detection of bit $j$. A zero value means that
the tested gain configuration is adapted to the bit sequence.}}\label{table}
\end{table}

The reading process to determine which bit sequence is lit on the disc follows these few steps :
\begin{itemize}
    \item the time dependent intensity is first measured on each of the five detectors with all electronic gains set to one.
    \item these intensities are integrated for a time $T$.
    \item the signal is then calculated, using the different gain configurations $j$
    \item the bit sequence effectively present in the focal spot is determined by the only signal yielding a zero value.
\end{itemize}
Note that the second step just corresponds to the $N_{k}$ measurements. The integration time $T$ is chosen as the time
interval during which the signal leads to the determination of a unique bit sequence. The third step corresponds to the
simple calculation of a line in table \ref{table}. This can be done in parallel thanks to the speed of data processing on
dedicated processors, and the reading rate will thus not be affected compared to current devices. Finally, note that the
last step requires a good choice of the parameters in order to be able to distinguish all bit sequences. It means that the
noise level has to be smaller than the difference between the two closest values from $0$, in order to get a zero mean
value for only one bit sequence. Indeed, there must be no overlap between the expectation values when we take into account
the noise and thus the uncertainty relative to each measurement. Note that using the zero value as the discriminating
factor could be combined with the use of all the calculated values, as each line of table (\ref{table}) is distinct. We
just need to know how to weight each data point according to the noise related to its obtention.

\section{Noise calculation}

\subsection{The shot noise limit}

To include the noise in our calculation, we separate classical and
quantum noise contributions. The classical noise comprises
residual noise of the laser diode, mechanical and thermal
vibrations. The major part of this noise is directly proportional
to the signal, i.e. to the number of detected photons. For a
detection of the total number of photons $N_{inc}$ in the whole
beam during the integration time of the detector, the classical
noise contribution $\sqrt{\langle\delta N_{inc}^{2}\rangle}$ would
thus be written as
\begin{equation}\label{noise1}
\sqrt{\langle\delta N_{inc}^{2}\rangle}=\beta N_{inc}
\end{equation}
where $\beta$ is a constant factor. And the individual noise
variable $\delta N_{i}(k)$ arising from detection on pixel $D_{k}$
is given by
\begin{equation}\label{noise2}
\delta N_{i}(k)= \frac{N_{i}(k)}{N_{inc}}\delta N_{inc}
\end{equation}
Using equations \ref{signal}, \ref{noise1} and \ref{noise2}, a
simple calculation yields the variance of the signal arising from
the classical noise
\begin{equation}
\langle \delta \hat{S}^{2}_{i}(j)\rangle_{Cl}=\frac{B
{S}^{2}_{i}(j)}{N_{inc}}
\end{equation}
where the constant $B=N_{inc}\beta^{2}$ is the classical noise
factor, and is chosen so that, when $B=1$ and when all the
intensity is detected by one detector, the classical noise term is
equal to the shot noise term.
Note that classical noise does not deteriorate measurements
having a zero mean value. For this reason, we have chosen to
discriminate bit sequences by choosing gains such as $S_{i}(i)=0$,
as mentioned earlier.

\begin{figure}[htbp]
\begin{center}
\includegraphics[width=8cm]{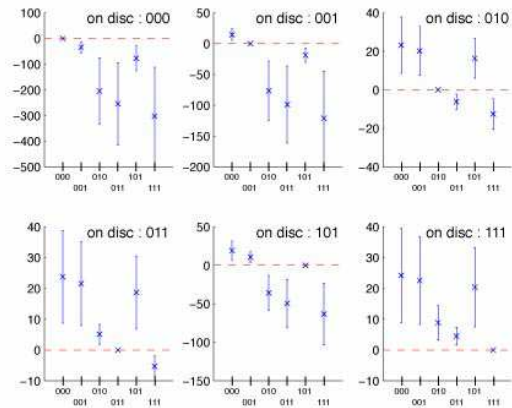}
\caption{Color 
online{\bf Classical noise ($10~dB$ of excess noise) represented
as error bars, for $\lambda=0.78 \mu m$, $NA=0.47$, and $25$
detected photons. Each inset corresponds to the $6$ signals
obtained for the different gain configurations, when one
particular bit sequence is present in the focal spot. Each bit
sequence present in the focal spot can be clearly identified as
only one gain configuration can give a zero value for each
inset.}} \label{classical}
\end{center}
\end{figure}
The calculation of the quantum contribution requires the use of
quantum field operators, describing the quantum fluctuations in
all transverse modes of the field. By changing the gain
configuration of the array detector, not only the signal
$S_{i}(j)$ is modified, but also the related quantum noise denoted
$\langle \delta\hat{S}^{2}_{i}(j)\rangle_{Qu}$, as different gain
configurations are sensitive to noise in different modes of the
field. We have shown in reference \cite{Treps} that for a
multi-pixel detection of an optical image, the measurement noise
arises from only one mode component of the field, referred to as
the {\it detection mode}, or {\it noise-mode} \cite{Del}. The
expression of the quantum noise is then :
\begin{eqnarray}\label{qnoise}
\langle \delta \hat{S}^{2}_{i}(j)\rangle_{Qu} = f_{i,j}^{2}
N_{inc}\langle\delta \hat{X}^{2}_{w_{i,j}}\rangle
\end{eqnarray}
where $\delta \hat{X}_{w_{i,j}}$ is the quantum noise contribution
of the noise-mode $w_{i,j}(x)$ which is defined for one set of
gain $j$, when the bit sequence $i$ is present in the focal spot,
as
\begin{equation}\label{wdef}
\forall x \in D_{k} \qquad w_{i,j}(x) =
\frac{\sigma_{k}(j)n_{i}(x)}{f_{i,j}}
\end{equation}
and where $f_{i,j}$ is a normalization factor, which expression is
\begin{equation}\label{fdef}
f^{2}_{i,j} =
\frac{\sum_{k=1}^{5}\sigma^{2}_{k}(j)N_{k}(i)dx}{N_{inc}}
\end{equation}
The noise-mode corresponds in fact to the incident field profile
weighted by the gains. The shot noise level corresponds to
$\langle\delta \hat{X}^{2}_{w_{i,j}}\rangle=1$.

The variance of the signal can eventually be written as:
\begin{eqnarray}\label{noisedef}
\langle \delta \hat{S}^{2}_{i}(j)\rangle = f_{i,j}^{2}
N_{inc}\langle\delta \hat{X}^{2}_{w_{i,j}}\rangle + \frac{B
{S}^{2}_{i}(j)}{N_{inc}}
\end{eqnarray}

\begin{figure}[htbp]
\begin{center}
\includegraphics[width=8cm]{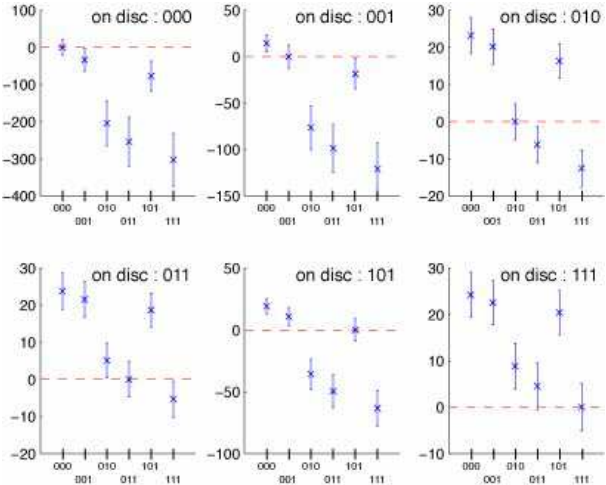}
\caption{Color 
online{\bf Shot noise represented as error bars, for
$\lambda=0.78 \mu m$, $NA=0.47$, $25$ detected photons. Some bit
sequences cannot be determined without ambiguity because of the
noise level.}} \label{shot}
\end{center}
\end{figure}
We have first represented the classical noise with an excess noise
of $10~dB$, as error bars for each result $S_{i}(j)$, on figures
(\ref{classical}). We have chosen a representation with a number
of detected photons of only $25$. Each of the $6$ insets refers to
the measurement obtained for a particular bit sequence in the
focal spot. The $6$ data points and associated error bars refer to
the results obtained when the $6$ gain configurations are tested.
One inset thus corresponds to one line in table (\ref{table}). We
can see that with this choice of parameters, the bit sequence
effectively present in the focal spot can be determined without
ambiguity by the only zero value. The sequence corresponds to the
one for which the gains were optimized. We see that bit sequence
discrimination can be achieved even with a very low number of
photons. The relative immunity to classical noise of our scheme
arises from the fact that measurements are performed around a zero
mean value. Thus, given this limit in the minimum necessary photon
number and the flux of photons one can calculate the maximum data
rate, which is found to be $2.10^{7} Mbits/s$ (this estimation
takes into account an integration time $T$ corresponding to $1/10$
of the delay between the read-out process of two adjacent bits
with a $1mW$ laser). This very high value shows that classical
noise should not be a limit for data rate in such a scheme.

The effect of quantum noise is very small, but becomes a limiting
factor for such a small number of detected photons, or for a large
number of bits encoded on the disc in the wavelength size. In
order to see independently the effect of each contribution to the
noise, we have thus represented on figure (\ref{shot}) the shot
noise also for $25$ detected photons, appearing as the threshold
under which it is impossible to distinguish bit sequences because
of the quantum noise. Note that for the represented case, the shot
noise is the most important contribution, and that it prevents a
bit sequence discrimination, as a zero value for the signal can be
obtained for several gain configurations in the same inset.

\subsection{Beyond the shot noise limit}

When the shot noise is the limiting factor, non classical light
can be used to perform measurements beyond the quantum noise
limit. We have shown in reference \cite{Treps} that squeezing the
noise-mode of the incident field was a necessary and sufficient
condition to a perfect measurement. What we are interested in is
improving the measurements that yield a zero value, which are
obtained when the gain configuration matches the bit sequence in
the focal spot, as $S_{i}(i)=0$. Using equation (\ref{noisedef}),
we see that $w_{i,i}$ has to be squeezed. As no information on the
bit present in the focal spot is available before the measurement,
in order to improve simultaneously all the bit sequences
detections, the $6$ noise-modes have to be squeezed at the same
time in the incident field. These $6$ transverse modes are not
necessarily orthogonal, but one can show that squeezing the
subspace that can generate all of them is enough to induce the
same amount of squeezing.
%
%
\begin{figure}[htbp]
\begin{center}
\includegraphics[width=8cm]{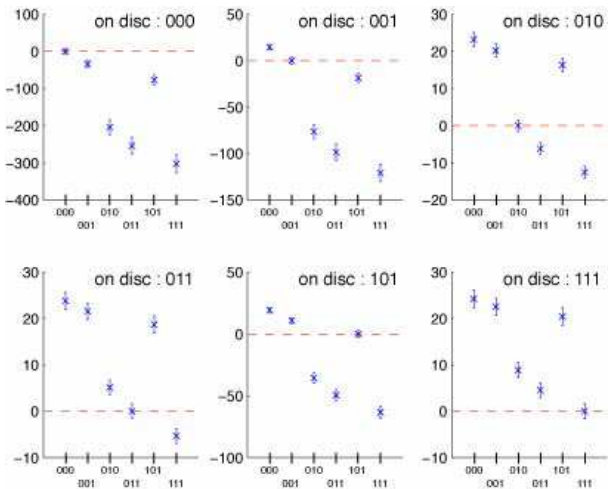}
\caption{Color 
online{\bf Quantum detection noise represented as error bars, for $\lambda=0.78 \mu m$, $NA=0.47$, $25$ detected
photons and $-10~dB$ of simultaneous squeezing for all the flipped modes. The ambiguity in presence of shot noise has been
removed and each bit sequence can be identified.}} \label{SQZ}
\end{center}
\end{figure}
%

The quantum noise with $10~dB$ of squeezing on the sub space generated by the $w_{i,i}$
is represented as error bars on figure (\ref{SQZ}). The noise
of each noise-mode $w_{i,j}$ is computed using its overlap integrals with the generator modes of the squeezed sub space, assuming that all modes orthogonal to the squeezed subspace are filled with coherent noise.
In this case, the effect of squeezing, reducing the quantum noise on the measurements, and
especially on the measurement for which the gains have been
optimized, is enough to discriminate bit sequences that were
masked by quantum noise.

\section{Conclusion}

We have proposed a novel way of information extraction from optical discs, based on multi-pixel detection. We have first
demonstrated, using only classical resources, that this detection could allow large data storage capacity, by burning
several bits in the spot size of the reading laser. We have presented a demonstration of principle through a simple
example which will be refined in further studies. We have also shown that in shot noise limited measurements, using
squeezed light in appropriate modes of the incident laser beam can lead to improvement in bit sequence discrimination.

The next steps are to study in details how to extract the redundant information when the disc is spinning, and to
systematically optimize the number of bits in the focal spot, the number and size of pixels in the array detector. Such a
regime involving a large number of bits in the focal spot will ultimately be limited by the shot noise, and will require
the quantum noise calculations presented in this paper.

\section{Acknowledgment}

We thank Magnus Hsu, Ping Koy Lam and Hans Bachor for fruitful
discussions.

Laboratoire Kastler Brossel, of the Ecole Normale Superieure and
University Pierre et Marie Curie, is associated to CNRS. This work
has been supported by the European Union in the frame of the
QUANTIM network (contract IST 2000-26019).

\end{document}